\documentclass[aip,reprint,amsmath,amssymb,superscriptaddress,showpacs,apl]{revtex4-1}
\usepackage{graphicx}
\usepackage{color}
\usepackage{lipsum}
\usepackage{dcolumn}

\begin{document}

\title{Percolation description of charge transport in the random barrier model applied to amorphous oxide semiconductors}

\author {S.~D.~Baranovskii}
\email {baranovs@staff.uni-marburg.de}
\affiliation{Department of Physics and Material Sciences Center,
Philipps-University, D-35032 Marburg, Germany}

\author{A.~V.~Nenashev}
\affiliation{Institute of Semiconductor Physics, 630090 Novosibirsk, Russia}
\affiliation{Novosibirsk State University, 630090 Novosibirsk, Russia}

\author {J.~O.~Oelerich}
\affiliation{Department of Physics and Material Sciences Center,
Philipps-University, D-35032 Marburg, Germany}

\author {S.~H.~M.~Greiner}
\affiliation{Department of Physics and Material Sciences Center,
Philipps-University, D-35032 Marburg, Germany}


\author{A.~V.~Dvurechenskii}
\affiliation{Institute of Semiconductor Physics, 630090 Novosibirsk, Russia}
\affiliation{Novosibirsk State University, 630090 Novosibirsk, Russia}

\author{F.~Gebhard}
\affiliation{Department of Physics and Material Sciences Center,
Philipps-University, D-35032 Marburg, Germany}



\date{\today}

\begin{abstract}
Charge transport in amorphous oxide semiconductors is often described as the band transport affected by disorder in the form of random potential barriers (RB). Theoretical studies in the framework of this approach neglected so far the percolation nature of the phenomenon. In this article, a recipe for theoretical description of charge transport in the RB model is formulated using percolation arguments. Comparison with the results published so far evidences the superiority of the percolation approach.


\end{abstract}







\maketitle   

Amorphous oxide semiconductors (AOSs) such as a-InGaZnO are in the focus of intensive research due to their applications in thin film transistors for transparent and flexible flat-panel displays \cite{Nomura2004_Nature,Kamiya2009,TAKAGI2005,Kamiya2010_APL,Kimura2010_APL}. Charge transport plays a decisive role in such applications. Therefore, much attention is currently dedicated to theoretical description of charge transport in AOSs.

Band transport via extended states affected by disorder potential is usually assumed as the dominant transport mechanism in AOSs. This assumption is based on experimental data that evidence a well-developed Hall effect implying band-like charge transport. Furthermore, the values of the charge carrier mobility $\mu > 10$ cm$^{2}$V$^{-1}$s$^{-1}$ measured in AOSs suggest band conduction as the dominant transport mechanism. These values are essentially larger than those expected for multiple trapping conduction or for hopping conduction \cite{Baranovski2006}, which could be considered as candidates for transport mechanism in AOSs \cite{Park2010,LeeParkKim2010,Germs2012}.

\begin{figure}[t]
\includegraphics[width=\linewidth]{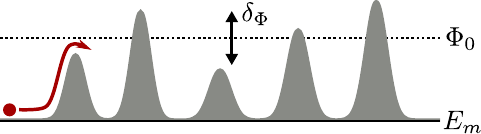}
\caption {Scheme of the model for band transport above the mobility edge $E_m$ affected by random potential barriers.  }
\label{fig:Kamiya_model}
\end{figure}

In their pioneering work, Kamiya \textit{et al}.\cite{Kamiya2009} assumed that charge carriers can move above the mobility edge $E_m$ where their motion is affected by disorder potential with Gaussian distribution
\begin{align}
  \label{DOS_Gauss}
  G_B(V) =
  \frac{1}{\delta_{\phi}\sqrt{2\pi}}
    \exp\left(-\frac{(V-\phi_0)
    ^{2}}{2{\delta_\phi}^{2}}\right) \, .
\end{align}
Here $\phi_0$ is the average height of the barriers and $\delta_{\phi}$ is the standard deviation in the distribution of the barrier heights. The model is sketched in Fig.~\ref{fig:Kamiya_model}.
Naming their study of charge transport in AOSs {\sl Percolation conduction examined by analytical model}\cite{Kamiya2009}, Kamiya \textit{et al.} ignored, however, the percolation nature of the conduction process. Instead, Kamiya \textit{et al}.\cite{Kamiya2009} based their consideration on averaging the transition rates for the activation of charge carriers over the potential barriers with the distribution of heights given by Eq.~(\ref{DOS_Gauss}).

The rate of the carrier activation over the barrier with height $V$ is equal to
\begin{equation}
\label{eq:actrates}
  \nu(V) =
 \nu_0 \exp\left(-\frac{eV}{kT}\right) \,
\end{equation}
where $e$ is the elementary charge and $\nu_0$ is the attempt-to-escape frequency. The latter parameter is of the order of the phonon frequency if transitions occur due to interaction of charge carriers with phonons. Averaging the activation rates given by Eq.~(\ref{eq:actrates}) over the distribution of barriers given by  Eq.~(\ref{DOS_Gauss}) yields the average rate $\langle\nu\rangle$
\begin{equation}
 \label{eq:averates}
\langle\nu\rangle =
 \nu_0 \exp\left[-\frac{e\phi_0}{kT} + \frac{(e\delta_{\phi})^2}{2(kT)^2} \right] ,
  \,
\end{equation}
which is dominated by barriers with the heights close to
\begin{equation}
\label{eq:averenergy}
V_{\nu} = \phi_0 - \frac{e \delta_{\phi}^2}{kT} \, .
\end{equation}
The same approach was also adopted by Lee \textit{et al}. \cite{Lee2011_APL}, who replaced the band mobility $\mu_0$ in the presence of disorder by
\begin{equation}
\label{eq:mutilda}
\mu_{\langle\nu\rangle}=\mu_0\exp\left[-\frac{e\phi_0}{kT} + \frac{(e\delta_{\phi})^2}{2(kT)^2} \right]
\end{equation}
calling the factor $\exp\left[-e\phi_0/(kT)+(e\delta_{\phi})^2/(2(kT)^2) \right]$
``a percolation term''.

In this work we will show that the above results are {\em not\/} related to percolation theory. First, the deficiencies of the rate averaging will be analyzed, and afterwards the recipe for description of charge transport via a system of random barriers (RB) will be formulated. Our percolation-theory description turns out to be in striking contrast to that based on the averaging of transition rates.

The irrelevance of the average rate $\langle\nu\rangle$ for characterizing charge transport in disordered systems with a broad distribution of microscopic transition rates, as given by Eq.~(\ref{eq:actrates}), has been recognized already in the early 1970s when the theoretical description of charge transport based on the percolation theory was developed \cite{Shklovskii1971,Ambegaokar1971,Pollak1972}. The deficiencies of an approach based on the averaging of transition rates $\langle\nu\rangle$  have been analyzed in detail in monographs \cite{Shklovskii1984}, edited books \cite{Baranovski2006} and topical reviews \cite{Baranovskii2014,Nenashev_Topical_2015}. Nevertheless, this particular approach is frequently used for charge transport in AOSs \cite{Kamiya2009,Lee2011_APL}. Therefore, it is instructive to analyze this approach once again and to show why this approach can hardly yield reliable results. The deficiencies of the rate averaging in the form of Eqs.~(\ref{eq:averates}) and (\ref{eq:mutilda}) are particularly transparent for one-dimensional (1d) transport. In 1d, carriers move via thermal activation over the barriers, being forced to climb over the barrier tops without an option to avoid the highest barriers. In contrast, in 3d, charge carriers avoid the activation to the top of the highest barriers, percolating between the barriers. The percolation nature of charge transport in the RB model was neglected in Eqs.~(\ref{eq:averates}) and (\ref{eq:mutilda}). Before addressing this topic, let us first analyse charge transport in 1d case.

In the RB model illustrated in Fig.~\ref{fig:Kamiya_model}, it is assumed that charge carriers can rapidly move through valleys at the mobility edge $E_m$ between the barriers and that this fast movement is interrupted by a slow activation of carriers over the barriers. In such an incoherent process, charge transport can be described using an illustrative electrotechnical analogy \cite{Shklovskii1984}. Each potential barrier can be viewed as a resistance, whose magnitude $R(V)$ is proportional to the time $\tau(V)=\nu^{-1}(V)$ necessary to overcome the potential barrier with the height $V$. Using Eq.~(\ref{eq:actrates}), one obtains
\begin{equation}
\label{eq:resistance}
  R(V) =
 R_0 \exp\left(\frac{eV}{kT}\right) \, ,
\end{equation}
where the prefactor $R_0$ is independent of the barrier height $V$. Then, the scheme of the RB model illustrated in Fig.~\ref{fig:Kamiya_model} can be considered in 1d case as a set of resistances connected in series.

In this picture, the activation rates $\nu(V)$ given by Eq.~(\ref{eq:actrates}) appear analogous to the conductances, i.e., to the inverse resistances $R^{-1}(V)$. Averaging the activation rates \cite{Kamiya2009,Lee2011_APL} is then analogous to the calculation of the resistivity of a series of resistances by averaging the conductances. This is surely incorrect, particularly in the case where the individual resistances in series have an exponentially broad distribution of magnitudes.  This broad distribution is governed by the exponential dependence of the resistances on the barrier heights given by Eq.~(\ref{eq:resistance}). In 1d, when percolation is not possible, the resistivity of a series of resistances is determined by the average resistance, i.e., by the average time for activation over the barriers
\begin{equation}
 \label{eq:avertimes}
\langle\tau\rangle =
 \nu_0^{-1} \exp\left[\frac{e\phi_0}{kT} + \frac{(e\delta_{\phi})^2}{2(kT)^2} \right] .
  \,
\end{equation}
The average time $\langle\tau\rangle$ is dominated by barriers with the heights close to
\begin{equation}
\label{eq:avertimeenergy}
V_{\tau} = \phi_0 + \frac{e \delta_{\phi}^2}{kT} \, .
\end{equation}
The mobility of charge carriers $\mu_{\langle\tau\rangle}$ is then proportional to the inverse of the average activation time $\langle\tau(V)\rangle^{-1}$,
\begin{equation}
\label{eq:mobavertimes}
\mu_{\langle\tau\rangle}=\mu_0\exp\left[-\frac{e\phi_0}{kT} - \frac{(e\delta_{\phi})^2}{2(kT)^2} \right] \, .
\end{equation}

It is worth to emphasize once again that the rate averaging \cite{Kamiya2009,Lee2011_APL}, which yields Eqs.~(\ref{eq:averates})--(\ref{eq:mutilda}) is equivalent to the calculation of the resistivity of a series of resistances by averaging the inverse resistances, i.e., the conductances. Due to the exponentially broad distribution of conductances prescribed by the exponential distribution of rates given by Eq.~(\ref{eq:actrates}), the average value is dominated by the very large conductances, i.e., by exponentially small resistances. It is apparent that such small resistances cannot be responsible for the resistivity of a system of resistances connected in series. Ascribing the dominant role to such small resistances leads to overestimating the mobility by the factor $\exp\left[ (e\delta_{\phi})^2/(kT)^2 \right]$.
The error caused by the rate averaging is particularly essential if $e\delta_{\phi} \gg kT$.

Another comment is necessary with respect to the averaging procedure used by Kamiya \textit{et al.}\cite{Kamiya2009}, adopted also by Lee \textit{et al.}\cite{Lee2011_APL}, that leads to Eqs.~(\ref{eq:averates})--(\ref{eq:mutilda}). This approach was borrowed from the study of the barrier inhomogeneities at Schottky contacts by Werner and G\"{u}ttler \cite{Werner1991}. In Schottky contacts, potential barriers can act parallel to each other along the area of a contact. Therefore, the electrotechnical analog is a system of resistances connected parallel to each other, where each barrier can be represented by an effective resistance determined by Eq.~(\ref{eq:resistance}). In order to calculate the resistivity of such a system, it is surely correct to average the inverse resistances, i.e., conductances represented by rates of carrier activation over the barriers. Therefore, the average activation rate given by Eq.~(\ref{eq:averates}) can be responsible for carrier injection through inhomogeneous Schottky contacts \cite{Werner1991}. This average rate has, however, nothing  to do with charge transport through a series of potential barriers illustrated in Fig.~\ref{fig:Kamiya_model}. For a series of barriers, not the average activation rate, but rather the average activation time given by Eq.~(\ref{eq:avertimes}) is the characteristic quantity responsible for charge transport. Hence, the carrier mobility is to be described by Eq.~(\ref{eq:mobavertimes}) and not by Eq.~(\ref{eq:mutilda}).

So far, we considered a 1d model, in which charge carriers have to overcome a series of barriers, being forced to become thermally activated to the barrier tops. Let us now consider a 3d case, in which carriers can move through the system of barriers avoiding the tops of the highest barriers. Percolation arguments will play a decisive role for description of charge transport in such a case.

In 3d, charge carriers are not obliged to climb to the tops of the highest energy barriers, being able to avoid such high barriers by percolating aside the highest barriers. The essence of
the percolation approach is that transport is determined not by
the average rates, or by the average times, but rather by the rates and times of those transitions
that are most difficult among the ones still relevant for long-range transport.
Conduction in disordered systems is in fact a percolation process, in which
charge transport is determined by the slowest transitions that are needed to provide a connected
path through the system.

In order to formulate a percolation criterion for charge transport in 3d via a system of random barriers, as applied to AOSs, let us modify slightly the model illustrated in Fig.~\ref{fig:Kamiya_model}. Let the volume of the material be occupied by cubic cells of two distinct types. One type of cells, called \textit{valleys}, does not contain potential barriers. Such cells provide for charge carriers a uniform energy level equal to the position of the mobility edge $E_m$ in the absence of barriers. The other type of cells, called \textit{barriers}, provide potential barriers
for charge carriers with the distribution of heights $V$ described by Eq.~(\ref{DOS_Gauss}). Let the volume fraction of valleys be $\xi$ and that of barriers $1-\xi$. Let us assume for simplicity that the volume of a single valley is equal to the volume of a single barrier and that valleys and barriers occupy cells on a 3d lattice grid. Then, the percolation problem in the RB  model can be mapped onto the site percolation problem on the corresponding lattice grid \cite{Shklovskii1984}. In such a system, the current can flow via sites arranged on a regular lattice grid. Some sites are blocked and they prevent transport to neighboring sites, while other sites are unblocked and they transfer current to the neighboring sites. Let the fraction of unblocked sites be $x$. The site percolation problem provides a solution $x_c$ for the minimal fraction of unblocked sites, that allow the current flow through the system over macroscopic distances. The value $x_c$, called \textit{percolation threshold}, depends on the particular structure of the lattice grid.

According to the percolation approach to charge transport in the RB model, one has to find the \textit{percolation level}, i.e., the minimal height of the potential barriers $V_c$, which the carriers still have to overcome in order to enable transport over macroscopic distances. Then the carrier mobility $\mu_{\textit{perc}}$
is determined as
\begin{equation}
\label{eq:mu_V_c}
\mu_{\textit{perc}} \simeq \mu_0\exp\left[-\frac{eV_c}{kT} \right] \, .
\end{equation}
In  Eq.~(\ref{eq:mu_V_c}), it is assumed that the prefactor $\mu_0$ for charge transport at the percolation level $V_c$ is equal to the band mobility at energy $E_m$.

It will be convenient for calculations to introduce a variable $\widetilde{V_c} \equiv V_c - \phi_0$, which  describes the position of the percolation level $V_c$ relative to the middle of the barrier distribution $\phi_0$.
For the Gaussian distribution of barrier heights given by Eq.~(\ref{DOS_Gauss}), the percolation level $\widetilde{V_c}$ is related to the percolation threshold of the site percolation problem $x_c$ via the equation
\begin{equation}
 \label{eq:percolationthreshold}
x_c = \xi + (1-\xi)\int^{\widetilde{V_c}/\delta_{\phi}}_{-\infty}\frac{1}{\sqrt{2\pi}}
    \exp\left(-\frac{t
    ^{2}}{2}\right)\textit{d}t \, .
\end{equation}

\begin{figure}[b]
\includegraphics[width=\linewidth]{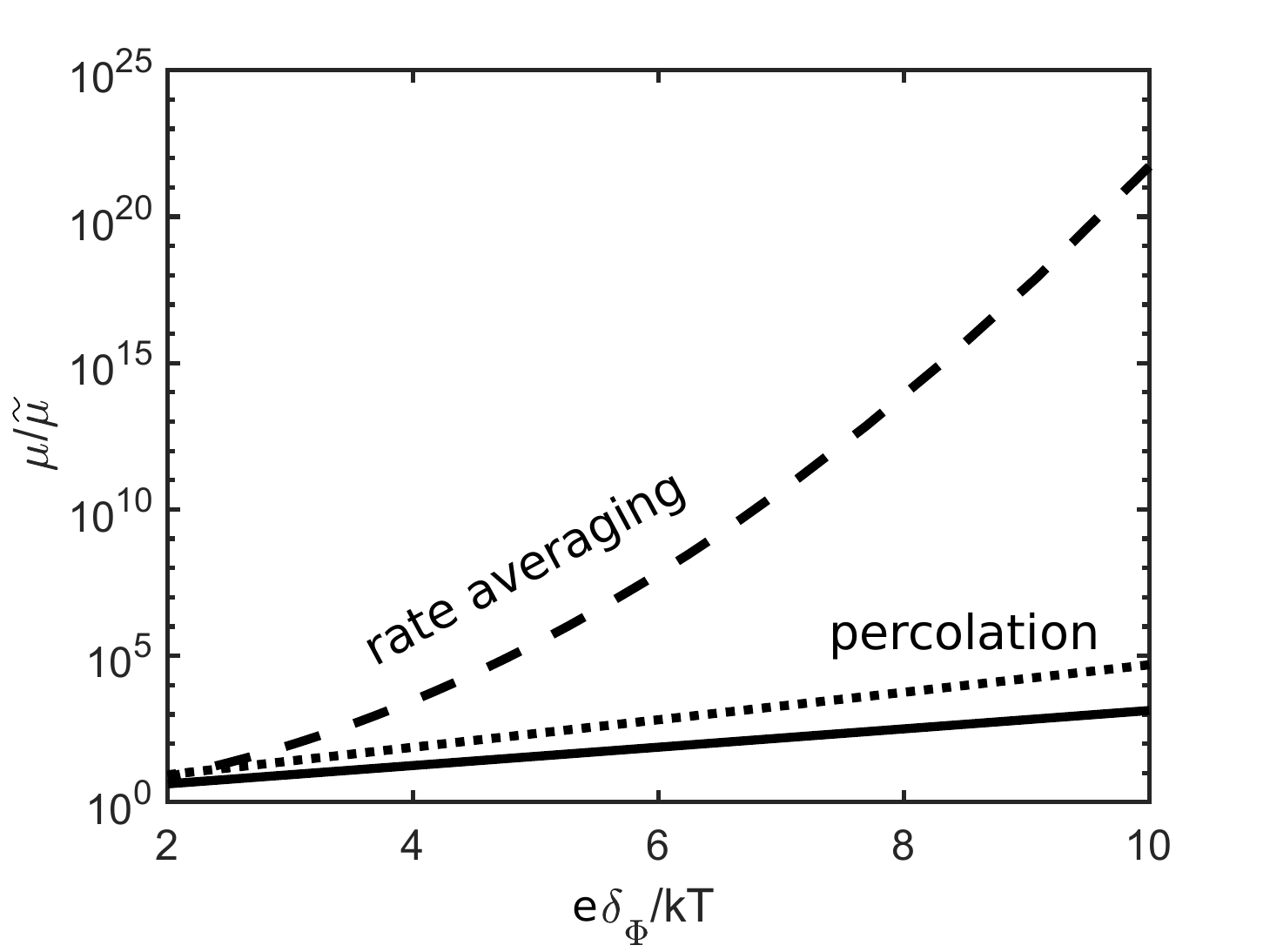}
\caption {Results of percolation theory (solid line for $\xi = 0.1$ and dotted line for $\xi = 0.2$) compared to the results of the rate averaging (dashed line).
  Plotted are $\mu_{\textit{perc}}$ given by Eqs.~(\ref{eq:percolationthreshold}) and~(\ref{eq:mupercRB}), and $\mu_{\langle\nu\rangle}$ given by Eq.~(\ref{eq:mutilda}),
  normalized by $\widetilde{\mu} = \mu_0\exp[-e\phi_0/(kT)]$.\label{fig:comparison}}
\end{figure}

Equation (\ref{eq:percolationthreshold}) yields the percolation level $\widetilde{V_c}$ as a function of the volume fraction of valleys $\xi$. If this fraction $\xi$ is larger than the percolation threshold $x_c$, no thermal activation is necessary for charge transport and carriers can move to macroscopic distances via spatially connected valleys at the energy level $E_m$. If $\xi < x_c$, percolation via valleys is interrupted by barriers, and thermal activation to the percolation level $V_c = \widetilde{V_c} + \phi_0$, where $\widetilde{V_c}$ is determined by Eq.~(\ref{eq:percolationthreshold}) becomes necessary. This yields the carrier mobility
\begin{equation}
\label{eq:mupercRB}
\mu_{\textit{perc}} \simeq \mu_0\exp\left[-\frac{e\phi_0}{kT} - \frac{e\widetilde{V_c}}{kT} \right] \, .
\end{equation}

In order to estimate the importance of percolation, we compare the results for carrier mobility given by Eq.~(\ref{eq:mupercRB}), with those given by Eq.~(\ref{eq:mutilda}).
The value of the percolation threshold $x_c$ is necessary for calculations of $\mu_{\textit{perc}}$ via Eqs.~(\ref{eq:percolationthreshold}) and (\ref{eq:mupercRB}). Let us assume for simplicity that valleys and barriers form a simple cubic lattice grid. For such a case one should use in Eq.~(\ref{eq:percolationthreshold}) the value \cite{Sykes1964,Skvor2009,Wang2013,Xu2014} $x_c\simeq 0.312$. At $\xi \geq 0.312$, barriers can be completely avoided by current flow at the level $E_m$. In order to bring barriers into play, $\xi$ should be smaller than $x_c$. The values $\xi = 0.1$ and $\xi = 0.2$ will be used in the calculations. Since both equations (\ref{eq:mupercRB}) and (\ref{eq:mutilda})
contain the factor $\widetilde{\mu} \equiv \mu_0\exp[-\frac{e\phi_0}{kT}]$, it is convenient to compare the ratios $\mu_{\textit{perc}} / \widetilde{\mu}$ and $\mu_{\langle\nu\rangle} / \widetilde{\mu}$.

In Fig.~\ref{fig:comparison}, the results of our percolation theory $\mu_{\textit{perc}} / \widetilde{\mu}$ expressed by Eqs.~(\ref{eq:percolationthreshold}) and (\ref{eq:mupercRB}) are plotted as functions of $e \delta_{\phi}/kT$ along with the result of the rate averaging \cite{Kamiya2009,Lee2011_APL}, $\mu_{\langle\nu\rangle} / \widetilde{\mu}$,
as expressed by Eq.~(\ref{eq:mutilda}).
In Fig.~\ref{fig:comparison}, the results of percolation theory are shown by a solid (dotted) line for $\xi=0.1$ ($\xi=0.2$), those from rate averaging are depicted by a dashed line.
  At low temperatures, compared to the width of the barrier distribution, $kT \ll e \delta_\phi$, the results from percolation theory differ by many orders of magnitude from the results
  given by the rate averaging. At high temperatures, when $kT$ is comparable with the width $e \delta_\phi$ of the barrier distribution, the distribution of the barrier heights does not play any role and the carrier mobility is close to the value
\begin{equation}
\label{eq:muno distribution}
\mu_{\textit{perc}} \simeq \mu_0\exp\left[-\frac{e\phi_0}{kT} \right] \,
\end{equation}
determined solely by the average barrier height $\phi_0$. In this case, taking the distribution of barrier heights into account is not necessary anyway.

The drastic difference between the results of the percolation theory expressed by Eqs.~(\ref{eq:percolationthreshold}) and~(\ref{eq:mupercRB}) as compared to those of the rate averaging expressed by Eq.~(\ref{eq:mutilda}) proves the importance of a percolation treatment of charge transport in the framework of the random barrier model,
widely used for electrical conduction in AOSs.




\begin{acknowledgments}

Financial support of the Deutsche Forschungsgemeinschaft (GRK 1782)
is gratefully acknowledged.
\end{acknowledgments}


\bibliography{Percolation}

\end{document}